\documentclass{PoS}%
\usepackage{amsmath}
\usepackage{amsfonts}
\usepackage{amssymb}
\usepackage{graphicx}%
\title{Quark-lepton flavored unification}
\ShortTitle{Flavored unification}
\author{\speaker{Christopher Smith}\\
Laboratoire de Physique Subatomique et de Cosmologie,
\linebreak
Universit\'{e} Grenoble-Alpes, CNRS/IN2P3, Grenoble INP, 38000 Grenoble, France\\
E-mail: \email{chsmith@lpsc.in2p3.fr}}

\abstract{The unification of the quark and charged lepton flavor structures is revisited using the tools of Minimal Flavor Violation (MFV). Under that framework, we first identify a unique point in parameter space where flavor universality is maximally violated. Though from the point of view of the usual polynomial MFV expansions, standing close to that point would require unacceptable fine-tunings, it is automatically attained with infinite geometric MFV series. Remarkably, this peculiar point is precisely where the charged lepton Yukawa coupling can be expressed naturally in terms of those of the quarks. The most striking consequence would be that the physical electron field is dominantly the gauge lepton field of the third generation.}

\FullConference{The 39th International Conference on High Energy Physics (ICHEP2018)\\
		4-11 July, 2018\\
		Seoul, Korea}

\begin{document}

\section{Introduction}

Not long after the advent of the Standard Model (SM), the possibility to further unify the fundamental interactions in some large but simple gauge group emerged. While this has proven very successful for the gauge sector, and is often regarded as one justification for the peculiar gauge quantum numbers of the SM fermions, these models ran into problems in the flavor sector. For example, the minimal SU(5) model has only two Yukawa couplings, and predicts $\mathbf{Y}_{d}=\mathbf{Y}_{e}^{T}$. Even accounting for QCD effects, this is in stark disagreement with measured quark and lepton masses. The standard way out is to introduce a third Yukawa coupling, in which case $\mathbf{Y}_{u,d,e}$ become uncorrelated and flavored unification is not achieved. In this talk, we report on the analysis of Ref.~\cite{Smith:2016ztg}, where the possibility of a relationship between the three SM Yukawa couplings was studied model-independently. Throughout this work, the neutrino sector is kept as trivial.

\section{Could the Yukawa couplings be redundant?}

To formulate the hypothesis of an intrinsic redundancy of the SM Yukawa couplings, we use the language of Minimal Flavor Violation (MFV)~\cite{DambrosioGIS02}. First, remember that gauge interactions being flavor blind, they are invariant under the flavor symmetry~\cite{FlavorSymm},
\begin{equation}
G_{F}=U(3)^{5}=U(3)_{q_{L}}\otimes U(3)_{u_{R}}\otimes U(3)_{d_{R}}\otimes U(3)_{\ell_{L}}\otimes U(3)_{e_{R}}\ ,
\end{equation}
with $q_{L}=(u,d)_{L}$, $\ell_{L}=(\nu_{e},e)_{L}$. What characterizes the flavor couplings is their breaking of $G_{F}$. Technically, they can be treated as spurions, i.e., non-dynamical fields carrying definite $G_{F}$ charges: $\mathbf{Y}_{u}\sim(\mathbf{\bar{3}},\mathbf{3},1,1,1)$,
$\mathbf{Y}_{d}\sim(\mathbf{\bar{3}},\mathbf{3},1,1,1)$, and $\mathbf{Y}_{e}\sim(1,1,1,\mathbf{\bar{3}},\mathbf{3})$.

Imagine now that at some high scale, there are only two fundamental flavor-breaking terms $\mathbf{Y}_{1}$ and $\mathbf{Y}_{2}$. At the low scale, $\mathbf{Y}_{u,d,e}$ must be induced by $\mathbf{Y}_{1,2}$, at which point they would be expressed as polynomials of the form $\mathbf{Y}_{u,d,e}=P_{u,d,e}(\mathbf{Y}_{1},\mathbf{Y}_{2})$. Provided these polynomials are
natural, i.e. involve $\mathcal{O}(1)$ coefficients, the unknown $\mathbf{Y}_{1,2}$ can be eliminated and one of the Yukawa coupling can be expressed as a natural polynomial in the other two. In practice, for this to work, we must choose a reduced flavor group and its two spurions. Inspired by the $SU(5)$ model, we take:%
\begin{equation}
G_{F}^{\prime}=U(3)^{3}=U(3)_{q_{L}=\ell_{L}}\otimes U(3)_{u_{R}}\otimes U(3)_{d_{R}=e_{R}}\ ,
\end{equation}
with $\mathbf{Y}_{1}\sim\mathbf{Y}_{u}\sim(\mathbf{\bar{3}},\mathbf{3},1)$ and $\mathbf{Y}_{2}\sim\mathbf{Y}_{d}\sim(\mathbf{\bar{3}},1,\mathbf{3})$. Then, $\mathbf{Y}_{e}$ is no longer fundamental but since it transforms as $\mathbf{Y}_{d}$, its general expression is a polynomial in $\mathbf{Y}_{u}$ and $\mathbf{Y}_{d}$ of the form
\begin{equation}
\mathbf{Y}_{e}=c_{0}\mathbf{Y}_{d}\cdot(\mathbf{1}+c_{1}\mathbf{Y}_{u}^{\dagger}\mathbf{Y}_{u}+c_{2}\mathbf{Y}_{d}^{\dagger}\mathbf{Y}_{d}+...)\ .\label{PolyExp}%
\end{equation}

Having established the $G_{F}^{\prime}$-invariant expansion of $\mathbf{Y}_{e}$, the spurions can now be frozen to their physical values. The $G_{F}^{\prime}$ symmetry is large enough to attain the gauge basis where $\mathbf{Y}_{u}\sim\mathbf{m}_{u}V_{CKM}$, $\mathbf{Y}_{d}\sim\mathbf{m}_{d}$.
Since $\mathbf{Y}_{e}$ is not necessarily diagonal in that basis, our goal is to find coefficients such that $g_{R}\mathbf{Y}_{e}g_{L}^{\dagger}\sim\mathbf{m}_{e}$ for some unitary $g_{R,L}$. Now, of course, this is always possible since $(\mathbf{Y}_{u}^{\dagger}\mathbf{Y}_{u})^{n}$ and$(\mathbf{Y}_{d}^{\dagger}\mathbf{Y}_{d})^{n}$ for $n=1,2,...$ form an algebraic basis for $3\times3$ matrices. What is crucial to the whole redundancy interpretation is that these coefficients must be of $\mathcal{O}(1)$. This is extremely constraining because the values of $\mathbf{Y}_{u}$
and $\mathbf{Y}_{d}$ are very peculiar, and the basis is nearly singular (the matrices $(\mathbf{Y}_{u,d}^{\dagger}\mathbf{Y}_{u,d})^{n}$ are nearly parallel to $\mathbf{Y}_{u,d}^{\dagger}\mathbf{Y}_{u,d}$)~\cite{ColangeloNS08}. To simplify the problem, let us keep only the first three terms in Eq.~(\ref{PolyExp}). Then, remarkably, the system does allow
for a natural solution in various scenarios~\cite{Smith:2016ztg}. For example, $c_{0,1,2}=20$, $-7.9$, $5.3$ in the MSSM assuming $\tan\beta=50$.

Looking back at Eq.~(\ref{PolyExp}), something special has to happen. If we write $\mathbf{Y}_{e}=c_{0}\mathbf{Y}_{d}\cdot\mathbf{X}$, the matrix $\mathbf{X}$ is responsible for transmuting the singular values of $\mathbf{Y}_{d}$ to that of $\mathbf{Y}_{e}$, which are significantly different. The only way this can happen with $c_{1,2}$ of $\mathcal{O}(1)$ is
to take advantage of the large top and bottom Yukawa couplings and ensure $1+c_{1}y_{t}^{2}+c_{2}y_{b}^{2}\approx0$, so that the matrix $\mathbf{X}$ is nearly singular with $\mathbf{X}^{33}\approx0\ll\mathbf{X}^{11,22}\approx1$. Interestingly, at that point, $g_{R}\mathbf{Y}_{e}g_{L}^{\dagger}\sim\mathbf{m}_{e}$ implies a reordering of the generations%
\begin{equation}
\left(
\begin{array}
[c]{c}%
e_{L}\\
\mu_{L}\\
\tau_{L}%
\end{array}
\right)  ^{gauge}\approx\left(
\begin{array}
[c]{ccc}%
0.03 & 0.98 & 0.20\\
0.06 & 0.20 & 0.98\\
1.00 & 0.02 & 0.07
\end{array}
\right)  \left(
\begin{array}
[c]{c}%
e_{L}\\
\mu_{L}\\
\tau_{L}%
\end{array}
\right)  ^{phys}\ .\label{Reorder}%
\end{equation}
In other words, the physical $e_{L}$ is actually the third generation leptonic
state, partnered with $(t,b)_{L}$.

\section{Geometric Minimal Flavor Violation}

\begin{figure}[t]
\centering\includegraphics[width=0.95\textwidth]{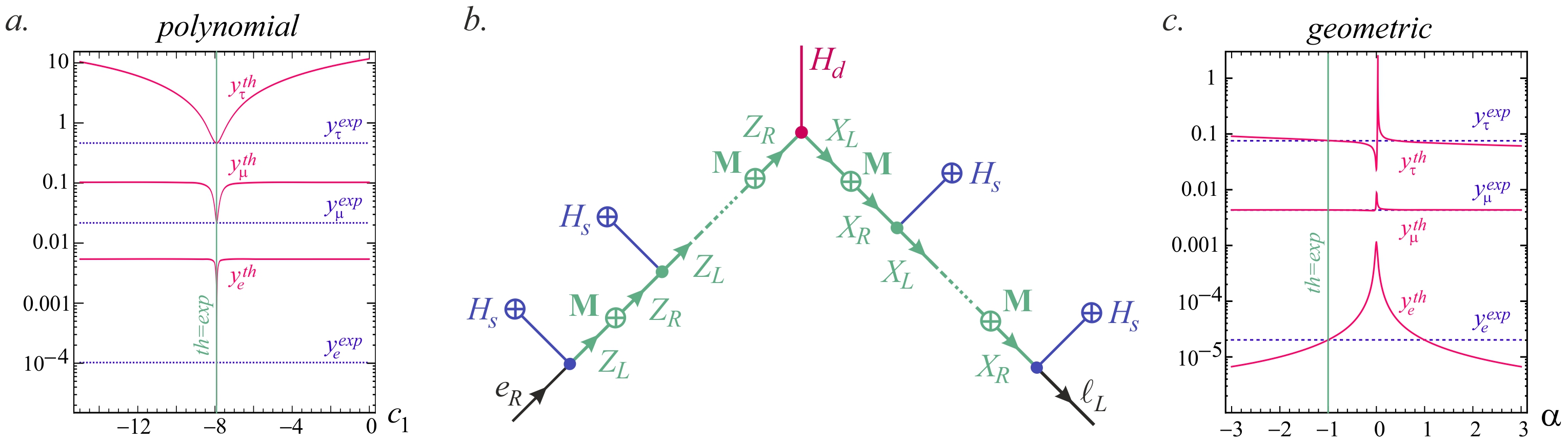}\caption{$a$)~Fine-tuning in the polynomial expansion Eq.~(\ref{PolyExp}). $b$)~Effective $\mathbf{Y}_{e}$ induced by heavy vector-like leptons. $c$)~Absence of fine-tuning in the geometric expansion Eq.~(\ref{GeomExp}). }%
\label{Fig}%
\end{figure}

Unfortunately, the solution found in the previous section is intrinsically unphysical. Indeed, as shown in Fig.~\ref{Fig}$a$, the singularity $\det\mathbf{X}=0$ is very narrow, so the $c_{i}$ have to be precisely fine-tuned to stay close to it. Adding terms to the expansion Eq.~(\ref{PolyExp}) does not change the picture, and we conclude that no finite polynomial relation between $\mathbf{Y}_{u}$, $\mathbf{Y}_{d}$, and $\mathbf{Y}_{e}$ will ever be truly natural.

Now, in general, there is no reason to expect the expansions to be finite. So, let us see if we can lift the fine-tuning by taking an infinite polynomial. Specifically, consider the geometric series:%
\begin{equation}
\mathbf{Y}_{e}(c_{i})=c_{0}\mathbf{Y}_{d}\cdot(\mathbf{1}
+\eta\mathbf{Y}_{u}^{\dagger}\mathbf{Y}_{u}
+\eta^{2}(\mathbf{Y}_{u}^{\dagger}\mathbf{Y}_{u})^{2}+...)
=c_{0}\mathbf{Y}_{d}\cdot(\mathbf{1}-\eta\mathbf{Y}_{u}^{\dagger}\mathbf{Y}_{u})^{-1}\ .
\end{equation}
The matrix $\mathbf{X}=(\mathbf{1}-\eta\mathbf{Y}_{u}^{\dagger}\mathbf{Y}_{u})^{-1}$ is diagonal with entries $(1-\eta y_{u,c,t}^{2})^{-1}$ in the basis where $\mathbf{Y}_{u}^{\dagger}\mathbf{Y}_{u}=\operatorname*{diag}(y_{u}^{2},y_{c}^{2},y_{t}^{2})$. So, for $\eta$ large enough, $\mathbf{X}\approx\operatorname*{diag}(1,1,0)$, which is precisely the structure we are after to solve $\mathbf{Y}_{e}=c_{0}\mathbf{Y}_{d}\cdot\mathbf{X}$. At this
stage, it may seem we have traded the fine-tuning of $\mathbf{X}$ for its convergence. Indeed, the fact that $(1-\eta y_{t})^{-1}\rightarrow0$ when $\eta y_{t}\gg1$ requires the series to be analytically continued. Thus, for this to make sense, the resummed series must be the true MFV prediction, in which case it is its expansion which is ill-defined when $\eta$ is too large.

To support this view, let us illustrate how imposing the usual polynomial MFV on some Lagrangian could lead to resummed geometric predictions. The idea is to view the geometric series as a Dyson series, with $\eta\mathbf{Y}_{u}^{\dagger}\mathbf{Y}_{u}$ playing the role of some mass insertion. Consider thus adding a set of weak doublet $X$ and weak singlet $Z$ vector-like leptons, transforming under $G_{F}^{\prime}$ as $\ell_{L}$ and $e_{R}$, respectively. Being vector like, these fields can be massive, and we assume $\mathbf{M}_{X}=\mathbf{M}_{Z}=M\mathbf{1}$ for simplicity. In addition to that, we introduce a gauge-singlet scalar field $H_{s}$ with flavored couplings to the leptons, on which we impose polynomial MFV. When $H_{s}$ acquire its vacuum expectation value $v_{s}$, the heavy lepton masses are shifted as $\delta\mathbf{M}_{X}=v_{s}(\alpha\mathbf{Y}_{u}^{\dagger}\mathbf{Y}_{u}+\beta\mathbf{Y}_{d}^{\dagger}\mathbf{Y}_{d})$ and $\delta\mathbf{M}_{Z}=v_{s}\varepsilon\mathbf{Y}_{d}\mathbf{Y}_{d}^{\dagger}$ for some $\mathcal{O}(1)$ coefficients $\alpha$, $\beta$, $\varepsilon$. Then, integrating out $X$ and $Z$ formally resums the infinite series shown in Fig.~\ref{Fig}$b$, and the total effective lepton Yukawa coupling takes the form
\begin{equation}
\mathbf{Y}_{e}=\left(  \mathbf{1}+\frac{v_{s}}{M}\varepsilon\mathbf{Y}_{d}\mathbf{Y}_{d}^{\dagger}\right)^{-1}
\cdot
\gamma\mathbf{Y}_{d}
\cdot
\left(  \mathbf{1}+\frac{v_{s}}{M}(\alpha\mathbf{Y}_{u}^{\dagger}\mathbf{Y}_{u}+\beta\mathbf{Y}_{d}^{\dagger}\mathbf{Y}_{d})\right)^{-1}\;,
\label{GeomExp}
\end{equation}
for some coefficient $\gamma$. Imposing $g_{R}\mathbf{Y}_{e}g_{L}^{\dagger}\sim\mathbf{m}_{e}$, infinitely many solutions exist, for example $\gamma=81$, $\alpha\equiv1$, $\beta\equiv1$, $\varepsilon\equiv-7$, $v_{s}/M=2.1\times10^{3}$. Importantly, no fine-tuning is implied, this solution is well-behaved (see Fig.~\ref{Fig}$c$), and pushing the geometric series out of its radius of convergence is accomplished thanks to the theoretically-free ratio $v_{s}/M$. We have thus achieved our goal of naturally generate $\mathbf{Y}_{e}$ out of $\mathbf{Y}_{u,d}$.\ The reordering observed in Eq.~(\ref{Reorder}) not only remains but is also extended to the right-handed leptons.

\section{Conclusion}

MFV is known to force all New Physics flavor couplings to depart only slightly from flavor universality (but for chirality-flip insertions whenever required), with all flavor mixings tuned by the CKM matrix. Yet, within the parameter range attainable with MFV, there is a peculiar and unique point where flavor universality can in principle be totally lifted. Expansions like $\mathbf{X}=\mathbf{1}+\alpha\mathbf{Y}_{u}^{\dagger}\mathbf{Y}_{u}$ become singular, $\det\mathbf{X}=0$, when $\alpha$ is close to $-\operatorname*{Tr}(\mathbf{Y}_{u}^{\dagger}\mathbf{Y}_{u})^{-1}$, which is a natural value since the top quark Yukawa coupling is of $\mathcal{O}(1)$. Coincidentally, setting $\alpha=-\operatorname*{Tr}(\mathbf{Y}_{u}^{\dagger}\mathbf{Y}_{u})^{-1}$ is not a fine-tuning but arises automatically if the MFV series is geometric in the mathematical sense. Remarkably, this unique singularity within the MFV range is precisely the one needed to unify the quark and lepton flavor structures. Clearly, the same reasoning can be extended to GUT settings.\ For example, in $SU(5)$, a successful unification could be obtained with only two fundamental Yukawa couplings. Extended to supersymmetric settings, geometric MFV structures generically predict light third generation squarks~\cite{NatSUSY}, and match natural SUSY-like scenarios. Implications for the neutrino sector remain to be studied, and could shed new light on the origin of their masses.

\end{document}